\documentclass[prb,preprint]{revtex4-1} 
\usepackage{amsmath}  % needed for \tfrac, \bmatrix, etc.
\usepackage{amsfonts} % needed for bold Greek, Fraktur, and blackboard bold
\usepackage{graphicx} % needed for figures

\begin{document}

% Be sure to use the \title, \author, \affiliation, and \abstract macros
% to format your title page.  Don't use lower-level macros to  manually
% adjust the fonts and centering.

\title{Riccati equations as a scale-relativistic \\gateway to quantum mechanics}
% In a long title you can use \\ to force a line break at a certain location.

\author{Saeed Naif Turki Al-Rashid}
\email{esp.saeedn.turkisntr2006@uoanbar.edu.iq} % optional
%\altaffiliation[permanent address: ]{101 Main Street, 
  %Anytown, USA} % optional second address
% If there were a second author at the same address, we would put another 
% \author{} statement here.  Don't combine multiple authors in a single
% \author statement.
\affiliation{Physics Department, College of Education for Pure Sciences, University Of Anbar, Ramadi, Iraq}
% Please provide a full mailing address here.

\author{Mohammed A.Z. Habeeb}
\email{mah@sc.nahrainuniv.edu.iq}
\affiliation{Department of Physics, College of Science,  Al-Nahrain University, Baghdad, Iraq}

\author{Stephan LeBohec}
\email{lebohec@physics.utah.edu}
\affiliation{Department of Physics and Astronomy, University of Utah, Salt Lake City, UT 84112-0830, USA}

% See the REVTeX documentation for more examples of author and affiliation lists.

\date{\today}

\begin{abstract}
Applying the resolution-scale relativity principle to develop a mechanics of non-differentiable dynamical paths, we find that, in one dimension, stationary motion corresponds to an It\^o process driven by the solutions of a Riccati equation. We verify that the corresponding Fokker-Planck equation is solved for a probability density corresponding to the squared modulus of the solution of the Schr\"odinger equation for the same problem. Inspired by the treatment of the one-dimensional case, we identify a generalization to time dependent problems in any number of dimensions. The It\^o process is then driven by a function which is identified as establishing the link between non-differentiable dynamics and standard quantum mechanics. This is the basis for the scale relativistic interpretation of standard quantum mechanics and, in the case of applications to chaotic systems, it leads us to identify quantum-like states as characterizing the entire system rather than the motion of its individual constituents.    
\end{abstract}
% AJP requires an abstract for all regular article submissions.
% Abstracts are optional for submissions to the "Notes and Discussions" section.

\maketitle % title page is now complete

%%------------------------------------------------------------------------------------------------------------------------
%%------------------------------------------------------------------------------------------------------------------------
\section{Introduction}
Scale relativity was proposed by Laurent Nottale\cite{nottale1993,nottale2007,nottale2011} to extend the relativity principle to transformations of resolution-scales, which become additional relative attributes defining reference frames with respect to one another. 
As such, scale relativity generalizes the standard theory of relativity and includes it as a special case, when the considered systems are invariant under changes of resolution-scale such as in classical mechanics with smooth differentiable trajectories. The {\it resolution-scale relativity principle} leads to the abandonment of the usually implicit hypothesis of differentiability and opens up on the consideration of non-differentiable or generally fractal geometries. This amounts to a clear departure from classical dynamics with, in particular, the loss of trajectories as a meaningful concept.

In this article, we concentrate at first on the stationary states, which, in standard quantum mechanics, are described as eigenfunctions of the Hamiltonian operator. We do not intend to provide a detailed review of the scale relativistic approach to standard quantum mechanics nor to discuss the details of its possible interpretations. More details can be found in publications by Nottale  \cite{nottale1993,nottale2007,nottale2011} or in an article by M.-H. Teh \cite{MHTeh2017}. However, in order to ensure some level of self-containment, in Section \ref{sr}, we provide the main lines of the application of the resolution-scale relativity principle for the development of non-differentiable dynamics. In particular, we show how concentrating on one dimensional stationary states leads to identifying an It\^o process driven by the solutions of a Riccati equation. This special form of Riccati equations commonly appears in quantum mechanics and, in Section \ref{ricsec}, after reviewing properties of Riccati equations, we see that finding their solutions entails solving the time independent Schr\"odinger equation for the same problem. In Section \ref{fp} we verify that, in the stationary regime, the Fokker-Planck equation for the It\^o process is satisfied by the squared modulus of the eigenfunctions of the time independent Schr\"odinger equation. In order to further expose the connection with standard quantum mechanics, in Section \ref{mad}, we travel the reverse route and show that the Schr\"odinger equation re-written as a Madelung system of equations\cite{madelung1927} yields the same Riccati equation in the one dimensional stationary case.  Then, in Section \ref{multidim}, we transpose these results to the multidimensional case and obtain a multidimensional It\^o process, the integration of which  yields standard quantum mechanics statistics also for time dependent states. This allows us to comment, in Section \ref{chaos}, on the possibility that the resolution-scale relativity principle is also applicable to chaotic or complex systems, which come under an effective stochastic description.  Finally, in Section \ref{conclusion}, we summarize and bring a few conclusions together.

%%------------------------------------------------------------------------------------------------------------------------
%%------------------------------------------------------------------------------------------------------------------------
\section{The Riccati equation from resolution-scale relativity}\label{sr}

This Section is inspired by Section 3, 4, and 5 in M.-H.\,Teh's paper \cite{MHTeh2017}, where more details and discussions can be found. In classical mechanics, the velocity is defined as the rate of change of the position ${\bf x}(t)$ during an infinitesimal time step 
forward or backward. The resolution-scale relativity principle leads to considering non-differentiable paths. This makes the velocity resolution-scale dependent as it can only be defined by considering finite time steps. Also, it breaks the symmetry between forward and backward time steps, resulting in the resolution-scale dependent  velocity to be double-valued. This situation can be dealt with by defining the following finite time differentials of position:
\begin{eqnarray*}
\frac{d_+}{dt}\bigg |_{\delta t}{\bf x}&=\frac{{\bf x}(t+\delta t,\delta t)-{\bf x}(t,\delta t)}{\delta t}&={\bf v}_+(t,\delta t){~~~~\delta t>0},\\
\frac{d_-}{dt}\bigg |_{\delta t}{\bf x}&=\frac{{\bf x}(t+\delta t,\delta t)-{\bf x}(t,\delta t)}{\delta t}&={\bf v}_-(t,\delta t){~~~~\delta t<0},
\end{eqnarray*}
with which a displacement over a time step $dt$ can be described by the combination of two terms: 
\begin{eqnarray}\label{repres}
d{\bf x}_+&=&{\bf v}_+dt+d{\bf b}_+{~~~~0<dt<\delta t},\nonumber \\
d{\bf x}_-&=&{\bf  v}_-dt+d{\bf b}_-{~~~~-\delta t<dt<0}.
\end{eqnarray}
In both expressions, the first term amounts to a displacement with a {\it usual velocity} ${\bf \dot x}_\pm$ depending on the resolution-scale $\delta t$. The second terms ${d\bf b}_\pm(t)$ correspond to possibly stochastic residuals such that $\langle {d\bf b}_\pm\rangle=0$ with the averaging being done over all the paths sharing the same usual velocities\cite{nottale2011}.  

Without any loss of generality, the two differential operators can be combined into a single {\it complex time-differential operator}\cite{nottale2011}: 
\begin{eqnarray}\label{complex_diff_op}
\frac{\hat d}{dt}=\frac{1}{2}\left(\frac{d_+}{dt}\bigg |_{\delta t}+\frac{d_-}{dt}\bigg |_{\delta t}\right)-\frac{i}{2}\left(\frac{d_+}{dt}\bigg |_{\delta t}-\frac{d_-}{dt}\bigg |_{\delta t}\right).
\end{eqnarray}
This can be used to define the {\it complex velocity}:
\begin{eqnarray}
\label{complexv}
\mathcal V=\frac{\hat d}{dt}{\bf x}=\frac{{\bf v}_++{\bf v}_-}{2}-i\frac{{\bf v}_+-{\bf v}_-}{2}={\bf V}-i{\bf U},
\end{eqnarray}
where $\bf V$ can be regarded as the classical velocity and $\bf U$ is an additional term, the {\it kink velocity}\cite{MHTeh2017}, which classically vanishes in the limit of infinitesimal $\delta t$ but persists and generally diverges for non-differentiable paths.

We now turn to considering a classical field $h({\bf x},t)$ on the Taylor expansion of which we act with the forward and backward time differential operators defined above and take the expectation value over all the compatible paths. This amounts to smoothing out all paths details smaller than some resolution-scale set by the inspection time-scale $\delta t$, which explicitly enters the definition of the differential operator:  
$$
\frac{d_\pm }{dt}h=\frac{\partial h}{\partial t}+\frac{\partial h}{\partial x_i}v_{i\pm}+\frac{\partial h}{\partial x_i}\langle\frac{db_{i\pm}}{dt}\rangle+\frac{1}{2}\frac{\partial^2 h}{\partial x_i\partial x_j}\langle\frac{db_{i\pm}db_{j\pm}}{dt}\rangle+\cdots
$$
If we now restrict ourselves to  $d\bf b_\pm$ being a Wiener process such that  $\langle db_{i+}\cdot db_{i-}\rangle=0$,  and $\langle db_{i+}\cdot db_{j+}\rangle=\langle db_{i-}\cdot db_{j-}\rangle=2\mathcal D\delta_{i,j}dt$, in the limit of infinitesimal times $dt$, the  complex and resolution-scale dependent time-differential of $h$ becomes\cite{nottale2011,MHTeh2017}:
\begin{eqnarray}\label{covardiff}
\frac{\hat d}{dt}h=\left[\frac{\partial }{\partial t}+\mathcal V\cdot \nabla -i \mathcal D\Delta \right]h.
\end{eqnarray}

This can be used to develop a mechanics of non-differentiable dynamical paths. For this purpose, we assume that mechanical systems can be characterized by a now complex Lagrange function $\mathcal L({\bf x},\mathcal V,t)$ and we correspondingly define the complex action as $\mathcal S=\int_{t_1}^{t_2}\mathcal L({\bf x},\mathcal V,t)dt$.  Enforcing the stationary action principle, while  keeping track of changes in the Leibniz product rule resulting from the higher order differential term in Equation \ref{complex_diff_op}, leads to the usual Euler-Lagrange equation but with the complex time-differential operator and velocity: 
$$
\frac{\partial \mathcal L}{\partial x}-\frac{\hat d}{dt}\left(\frac{\partial \mathcal L}{\partial \mathcal V}\right)=0.
$$

We then assume that the Lagrange function for a point particle of mass $m$ under the influence of a real potential energy term $\Phi$ keeps its classical form with the complex velocity in the place of the usual velocity:  $\mathcal L=\frac{1}{2}m\mathcal V^2-\Phi$. The Euler-Lagrange equation leads to  a generalized form of Newton's relation of dynamics where the velocity is replaced by the {\it complex velocity} (Equation \ref{complexv}) and the time derivative is replaced by the {\it complex finite time differential operator}  (Equation \ref{covardiff}):
\begin{eqnarray}\label{newton}
m\frac{\hat d}{dt}\mathcal V=-\nabla\Phi.
\end{eqnarray}
We may now replace $\mathcal V$ and $\frac{\hat d}{dt}$  by their expressions (Equations \ref{complexv} and \ref{covardiff}), so as to separate the real and imaginary parts of the generalized Newton Equation \ref{newton}:  
\begin{equation}
\begin{aligned}
\frac{\partial}{\partial t}{\bf V}-\mathcal D\Delta{\bf U}+({\bf V}\cdot\nabla){\bf V}-({\bf U}\cdot\nabla){\bf U}&=&-\frac{1}{m}\nabla\Phi,\\
\frac{\partial}{\partial t}{\bf U}+\mathcal D\Delta{\bf V}+({\bf V}\cdot\nabla){\bf U}+({\bf U}\cdot\nabla){\bf V}&=&0.
\end{aligned}
\label{nelson}
\end{equation}
It should be noted that this system of equations was also obtained by E.\,Nelson \cite{Nelson1966} in the context of stochastic mechanics with an entirely different set of hypothesis and interpretations\cite{nottale2011,MHTeh2017}. It can be noted then when there is no kink velocity ${\bf U}=0$, then standard classical mechanics is recovered. Inversely, we are now going to concentrate on a form of {\it stationary motion}, in which there is no usual velocity or drift but only the stochastic fluctuation. This amounts to setting $\langle{\bf V}\rangle=0$ and, in order to simplify, we restrict ourselves to the special case ${\bf V}=0$. The motion that is left is entirely described by the kink velocity {\bf U}, which is associated with the non-differentiable nature of paths that are otherwise stationary. Under this restriction, the system of Equations \ref{nelson} becomes:   
\begin{eqnarray}
\begin{aligned}
\mathcal D\Delta{\bf U}+({\bf U}\cdot\nabla){\bf U}&=\frac{1}{m}\nabla\Phi,\\
\frac{\partial}{\partial t}{\bf U}&=0.
\end{aligned}
\label{nelsonstation}
\end{eqnarray}
The second equation indicates that, as can be expected, $\bf U$ does not depend on time.  Since ${\bf V}=0$, we have ${{\bf v}_+=-{\bf v}_-}$, and ${\bf v}_+={\bf U}$. Consequently, Equation \ref{repres} becomes
\begin{equation}
\label{langevin}
d{\bf x}_+={\bf U}({\bf x})dt+d{\bf b}_+, 
\end{equation}
which is a Langevin equation or It\^o process driven by $\bf{U}({\bf x})$, solution to Equation \ref{nelsonstation} and in which
 $d{\bf b}_+$ is a stochastic function such that $\langle d{\bf b}_+ \rangle=0$ and $\langle db_{i+}\cdot db_{j+}\rangle=2\mathcal D\delta_{i,j}dt$.

We now further restrict ourselves and consider one dimensional problems for which Equation \ref{nelsonstation} becomes
\begin{equation}\nonumber
\frac{d}{dx}\left(\mathcal D\frac{dU}{dx}+\frac 12 U^2  \right)=\frac{1}{m}\frac{d\Phi(x)}{dx},
\end{equation}
which, when integrated once, takes the form of a Riccati equation\cite{bonilla2017}:
\begin{equation}\label{nelsonriccati}
U'=\frac{1}{m\mathcal D}\left(\Phi(x)-E\right)-\frac {1}{2\mathcal D} U^2, 
\end{equation}
 where $E$ is an integration constant with the dimension of energy. Here, we start using the lighter notation $U'=\frac{dU}{dx}$. With the resolution-scale relativity principle leading to standard quantum mechanics\cite{nottale2011,MHTeh2017} , it is not surprising to see the appearance of this Riccati equation. Several authors  \cite{haley1997,wheeler2004} investigated the occasional possibility of using Riccati equations to solve quantum mechanical problems such as a particle in a box, the simple harmonic oscillator and others. More fundamentally, G.W. Rogers \cite{rogers1985} observed that  the one-dimensional Schr\"odinger equation can be reduced to a Riccati form. This is particularly interesting: while being non-linear, the Riccati equation can be used to formulate standard quantum mechanics \cite{schuch2014}, which is generally presented as a fundamentally linear theory. Here, we identify again the Riccati equation as being deeply rooted in the quantum mechanical behavior but, this time, in a stochastic description, which independently manifests itself in the resolution-scale relativity framework without invoking quantum mechanics.
%%------------------------------------------------------------------------------------------------------------------------
%%------------------------------------------------------------------------------------------------------------------------
\section{Riccati equations}\label{ricsec}
A Riccati equation is a first order non-linear differential equation quadratic in the unknown function:
\begin{equation}\nonumber
U'(x)=q_0(x)+q_1(x)U(x)+q_2(x)U^2(x).
\end{equation}
Equation \ref{nelsonriccati} is of this type with $q_0=\frac{1}{m\mathcal D}\left(\Phi(x)-E\right)$, $q_1=0$, and $q_2=-\frac {1}{2\mathcal D}$.
With their relatively simple form, these equations constitute an attractive gateway to non linear dynamical systems. M. Nowakowski and H. C. Rosu \cite{NowakowskiRosu2002} investigated systems for which the equations of motion resulting from Newton's laws can be written as Riccati equations.  We have seen above that quantum mechanics can sometimes be formulated in terms of Riccati equations. These equations then appear as a possible point of connection between nonlinear dynamics and quantum mechanics.

An interesting property of the general Riccati equation \cite{reid1972} is that it can be rewritten as a linear second order differential equation by noticing that with $\eta=Uq_2$,  $$\eta'=q_0q_2+\left(q_1+\frac{q_2'}{q_2}\right)\eta+\eta^2,$$
%\begin{eqnarray}
%\eta'&=&U'q_2+Uq_2'\nonumber\\
%&=&(q_0+q_1U+q_2U^2)q_2+\eta\frac{q_2'}{q_2}\nonumber\\
%&=&q_0q_2+\left(q_1+\frac{q_2'}{q_2}\right)\eta+\eta^2\nonumber
%\end{eqnarray}
in which, substituting $\eta=-\frac{\psi'}{\psi}$ leads to $$-\psi''=q_0q_2\psi-\left(q_1+\frac{q_2'}{q_2}\right)\psi'.$$
%\begin{eqnarray}
%\frac{\psi'^2-\psi''\psi}{\psi^2}&=&q_0q_2-\left(q_1+\frac{q_2'}{q_2}\right)\frac{\psi'}{\psi}+\frac{\psi'^2}{\psi^2}\nonumber\\
%\psi'^2-\psi''\psi&=&q_0q_2\psi^2-\left(q_1+\frac{q_2'}{q_2}\right)\psi\psi'+\psi'^2\nonumber\\
%-\psi''&=&q_0q_2\psi-\left(q_1+\frac{q_2'}{q_2}\right)\psi'\nonumber
%\end{eqnarray}
When the solution $\psi$ of this equation is found, the solution of the original Riccati equation is obtained as $U=-\frac{\psi'}{q_2 \psi}$. 
In the case of Equation \ref{nelsonriccati}, $q_1+\frac{q_2'}{q_2}=0$ and the corresponding linear second order differential equation can be rearranged in the form
\begin{equation}\label{schrodinger}
-2m\mathcal D^2 \psi''+\Phi(x)\psi=E\psi,
\end{equation}
which, with the substitution  $\hbar\leftrightarrow 2m\mathcal D$, is just the time independent Schr\"odinger equation for a particle of mass $m$ in a one dimensional potential $\Phi$. This is the standard quantum mechanical answer to the problem we approached by applying the generalized Newton fundamental relation of dynamics Equation \ref{newton}. So the connection with standard quantum mechanics is starting to reveal itself. 

With $q_2=-\frac{1}{2\mathcal D}$, we have $U=2\mathcal D \frac{\psi'}{\psi}$ and with $\mathcal D>0$, the sign of $U$ is such that, in Equation \ref{langevin}, it corresponds to a flux toward regions where the wave function $\psi$ reaches extrema and away from its nodes. In the stationary case, this flux must be statistically compensated for by the stochastic process as can be verified with the Fokker-Planck equation.  

%%------------------------------------------------------------------------------------------------------------------------
%%------------------------------------------------------------------------------------------------------------------------
\section{Fokker-Planck equation}\label{fp}
Several authors \cite{MHTeh2017,AlRashid2006,AlRashid2007,AlRashid2011,Hermann1997} have applied the scale relativity approach to various time independent one-dimensional quantum problems. Before them, McClendon and Rabitz\cite{McClendon1988} applied the same Equations \ref{nelson} and \ref{langevin} but derived from Nelson's approach\cite{Nelson1966}. All verified that the numerical integration of the Langevin equation or It\^o process, Equation \ref{langevin}, yields position distributions converging toward the probability densities matching the stationary wave-functions given by standard quantum mechanics for the same potential. Here, we can see why this is so in all cases. The stochastic process described by Equation \ref{langevin} with $db=\sqrt{2\mathcal D}dW$, where $dW$ is the standard Wiener process,  corresponds to the following Fokker-Planck differential equation for the probability density $\rho$:
\begin{equation}
\frac{\partial \rho}{\partial t}=-\frac{\partial}{\partial x}\left(\rho U\right)+\mathcal D\frac{\partial^2\rho}{\partial x^2}.
\end{equation}
We just established that $U=2\mathcal D \frac {\psi'} {\psi}$ with $\psi$ a solution of the time independent Schr\"odinger Equation \ref{schrodinger}. In an attempt to identify the probability $\rho$ with the squared modulus of a probability amplitude, we may write the wave function as $\psi=\sqrt{\rho}e^{i\chi}$ and then $\psi'=\left(\frac{\rho'}{2\sqrt{\rho}}+i\sqrt\rho\chi'\right)e^{i\chi}$ and, in one dimension, in the stationary regime $\chi'=0$, so $U=\mathcal D \frac {\rho'} {\rho}$ and, hence, $\rho=\psi\cdot \psi^*$ solves the Fokker-Planck equation for stationarity $\frac{d\rho}{dt}=0$. 

This confirms that in the limit of infinite times, the integration of the It\^o process of Equation \ref{langevin} reproduces the quantum probability density for the position of a particle of mass $m$  in a stationary state of energy $E$ in any one dimensional potential $\Phi$. So, the connection with standard quantum mechanics is now explicit. 
%%------------------------------------------------------------------------------------------------------------------------
%%------------------------------------------------------------------------------------------------------------------------
\section{Madelung's equations}\label{mad}
To be complete, we can now start from standard quantum mechanics and see if we can identify the It\^o process that reproduces stationary quantum statistics. Writing the wave-function $\psi=\sqrt{\rho}e^{i\chi}$ in the usual definition of the probability current density ${\bf J}=\rho{\bf V}=-i\mathcal D\left(\psi^*\nabla\psi-\psi\nabla\psi^*\right)$, we identify the drift velocity as ${\bf V}=2\mathcal D\nabla\chi$. Similarly, 
Schr\"odinger's Equation \ref{schrodinger} can be rewritten as the equivalent system of Madelung's equations \cite{madelung1927}: 
\begin{eqnarray}
{{\partial \rho}\over{\partial t}}&=&-\nabla\left(\rho {\bf V}\right)\label{madelungcont},\\
({{\partial}\over{\partial t}}+{\bf V}\nabla){\bf V}&=&-{{\nabla(\Phi+\mathcal Q)}\over{m}}.\label{madelungeuler}
\end{eqnarray}

The first equation, Equation \ref{madelungcont}, is a continuity equation and the second, Equation \ref{madelungeuler}, is Euler's equation of fluid dynamics with $\mathcal Q=-2m\mathcal D^2{{\Delta\sqrt{\rho}}\over{\sqrt{\rho}}}$ as an additional term known as the quantum potential \cite{bohm1952} that is entirely responsible for the quantum behavior. 
In the one-dimensional stationary case discussed earlier, $\frac{\partial \rho}{\partial t}=0$ and also $V=0$, so the continuity equation is degenerate and, after integration, the Madelung-Euler fluid dynamics equation becomes $\mathcal Q=E-\Phi$, where $E$ is an integration constant. Substituting $\mathcal Q$ with its expression and expanding leads to
%\begin{equation}
%-2m\mathcal D^2{\frac{\partial^2\sqrt{\rho}}{\partial x^2}}=\left(E-\Phi\right)\sqrt{\rho}
%\end{equation}
%\begin{equation}
%-2m\mathcal D^2\frac{\partial}{\partial x}\left(\frac{\rho'}{2\sqrt{\rho}} \right)=\left(E-\Phi\right)\sqrt{\rho}
%\end{equation}
%\begin{equation}
%-m\mathcal D^2\frac{\rho''\sqrt{\rho}-\frac{\rho'^2}{2\sqrt{\rho}}}{\rho}=\left(E-\Phi\right)\sqrt{\rho}
%\end{equation}
%\begin{equation}
%-m\mathcal D^2\frac{\rho''\rho-\frac{\rho'^2}{2}}{\rho}=\left(E-\Phi\right)\rho
%\end{equation}
%\begin{equation}
%-m\mathcal D^2\left(\rho''\rho-\frac{\rho'^2}{2}\right)=\left(E-\Phi\right)\rho^2
%\end{equation}
\begin{equation}
-m\mathcal D^2\left(\frac{\rho''}{\rho}-\frac{1}{2}\left(\frac{\rho'}{\rho}\right)^2\right)=E-\Phi\nonumber.
\end{equation}
%\begin{equation}
%-m\mathcal D^2\left(\frac{\rho''}{\rho}-\frac{U^2}{2\mathcal D^2}\right)=\left(E-\Phi\right)
%\end{equation}
From this,  noting that $\frac{\rho''}{\rho}=\left(\frac{\rho'}{\rho}\right)'+\left(\frac{\rho'}{\rho}\right)^2$ and using $U=\mathcal D \frac {\rho'} {\rho}$ we obtain again the Riccati Equation \ref{nelsonriccati}. This time, however, it is obtained from a standard quantum mechanical  approach. The solutions of this Riccati equation retain the properties established earlier of driving the It\^o process of Equation \ref{langevin} in a way that reproduces stationary quantum probability densities. This convergence of Schr\"odinger's Equation \ref{schrodinger} and Newton's generalized Equation \ref{newton} of dynamics does not come as a surprise as, as further discussed in the next section, the two are in fact equivalent\cite{nottale2011,MHTeh2017}.  We now turn to a generalization of all this in more than one dimension. 
%%------------------------------------------------------------------------------------------------------------------------
%%------------------------------------------------------------------------------------------------------------------------
\section{More than one dimension}\label{multidim}
In more than one dimension, energy eigenfunctions cannot always be real and the complex velocity $\mathcal V$ generally has both a real and an imaginary part. We have seen above that in one dimension, for a state $\psi$, the kink velocity $U=2\mathcal D\frac{\psi'}{\psi}$ sets the flux required to compensate the diffusion resulting from the stochastic term in the It\^o process described by Equation \ref{langevin}. In more than one dimension, the kink velocity $\bf U$ should retain this role as it is the non-classical or {\it quantum-like} part of the complex velocity field. In addition, there can be a non-zero drift velocity $\bf V$ which is the real and classical part of the complex velocity field. Considering the expression for the complex velocity $\mathcal V={\bf V}-i{\bf U}$, in order to generalize the expression $U=2\mathcal D\frac{\psi'}{\psi}$, it is tempting, as an hypothesis to write $\mathcal V$ in the form:  \begin{equation}\mathcal V=-2i\mathcal D\frac{\nabla \psi}{\psi}.\label{3dvito}\end{equation}
Proceeding with $\psi=\sqrt{\rho}e^{i\chi}$ as before, we obtain
%$$\mathcal V=-\frac{2i\mathcal D}{\sqrt{\rho}e^{i\chi}}\left(i\sqrt{\rho}\nabla\chi+\frac{\nabla\rho}{2\sqrt{\rho}}\right)e^{i\chi}$$
$$\mathcal V=2\mathcal D\nabla\chi-i\mathcal D\frac{\nabla\rho}{\rho},$$
from which we identify ${{\bf V}=2\mathcal D \nabla\chi}$, which is the familiar expression for the drift velocity associated with the probability current density, and ${\bf U}=\mathcal D\frac{\nabla \rho}{\rho}$. The one dimensional result $U=2\mathcal D\frac{\psi'}{\psi}$ obtained by solving the Riccati Equation \ref{nelsonriccati} appears as an accidental consequence of the constancy of the complex argument of the wave function or absence of probability current. The expression $U=\mathcal D \frac {\rho'} {\rho}$ found in Section \ref{fp} now appears as more fundamental and general.

It is interesting to see how this works in the three dimensional Fokker-Planck equation with now $\bf V\ne 0$. With the definition of $\bf V$ and $\bf U$ in Equation \ref{complexv}, the time forward Equation \ref{repres} becomes \begin{equation}d{\bf x}_+=\left({\bf V+ U}\right)dt+d{\bf b}_+\label{ito3d}\end{equation} and the corresponding Fokker-Planck equation is 
$$
\frac{\partial \rho}{\partial t}=-\nabla\cdot\left(\rho {\bf V}\right)-\nabla\cdot\left(\rho {\bf U}\right)+\mathcal D\nabla^2\rho.
$$
With ${\bf U}=\mathcal D\frac{\nabla \rho}{\rho}$, the second term on the right-hand side is cancelled out by the third. The first term and the left-hand side constitute together Madelung's continuity Equation \ref{madelungcont}.  So, even in the time dependent cases, this Fokker-Planck equation is satisfied by our hypothesis for the complex velocity $\mathcal V$ with the real part corresponding to the drift associated with the probability current density and the imaginary part  corresponding to an {\it anti-diffusion}\cite{nottale2009} flux statistically compensated for by the stochastic process as already identified in the one dimensional case. 

%$$
%\frac{\partial \rho}{\partial t}=-\nabla\rho\cdot V-\rho\nabla\cdot V. 
%$$
%$$
%\frac{\partial \rho}{\partial t}=-2\mathcal D\nabla\rho\cdot \nabla\chi-2\mathcal D\rho\nabla^2\chi. 
%$$

In the one dimensional case, starting from the Madelung-Euler Equation \ref{madelungeuler} and using  $U=\mathcal D \frac {\rho'} {\rho}$, we obtained the Riccati equation, which we had previously obtained from the integration of the one dimensional version of Equation \ref{nelson} in the case $\bf V=0$. In the more than one dimension case, with the hypothesis we made for the expression of the complex velocity $\mathcal V$, we need to verify if the Madelung-Euler Equation \ref{madelungeuler} can again be connected with Equation \ref{nelson}. Or instead, we could equivalently connect Schr\"odinger's Equation \ref{schrodinger} and the generalized Equation \ref{newton} of dynamics. We do not reproduce the development\cite{nottale2011,MHTeh2017} here but the second can actually be shown to be equivalent to the first by writing the complex action in logarithmic form $\mathcal S=-2im\mathcal D\ln\psi$ (the constant $2m\mathcal D$ plays the role of a unit of action), which makes the complex velocity canonically appear as $\mathcal V=-2i\mathcal D \nabla \ln\psi$, which is precisely the complex velocity expression hypothesized in Equation \ref{3dvito}. 

This establishes that, with the complex velocity of Equation \ref{3dvito}, the integration of the It\^o process of Equation \ref{ito3d} statistically reproduces the quantum probability density for a time dependent state $\psi$ in any number of dimensions. 

%%------------------------------------------------------------------------------------------------------------------------
%%------------------------------------------------------------------------------------------------------------------------
\section{On quantum-like mechanics and dynamical chaos}\label{chaos}
We have seen above that enforcing the resolution-scale relativity principle to conservative point mechanics problems leads to an It\^o process (Equation \ref{ito3d}) driven by a complex velocity, which can be written as $\mathcal V={\bf V}-i{\bf U}=-2i\mathcal D \nabla \ln\psi$ with $\psi$ the solution of the Schr\"odinger equation for the same problem. Integration of this It\^o process then statistically reproduce the position probability density $|\psi|^2$. One specific integration of the It\^o process amounts to sampling one of an infinite number of dynamical paths. It should be clear that such a sampling is not to be identified with the state of the system in any way. Under the resolution-scale relativity principle, when the dynamics is resolution-scale dependent with a Wiener process component, notions of position and trajectory lose their meaning. The paths are neither enumerable nor distinguishable, so the system can not be described as following a particular one. The sampling of dynamical paths obtained by integration of the It\^o process does not correspond to any physical reality as long as the successive positions are not given by successive position measurements. Instead, the state of the system is to be identified with a time section of the entire bundle of dynamical paths\cite{nottale2011,MHTeh2017} with the same resolution-scale dependent complex velocity $\mathcal V$, which was found to correspond to the gradient of the logarithm of the wave function used to describe the state of the system in standard quantum mechanics. 

It has been suggested that if the resolution-scale relativity principle is applicable to complex or chaotic systems that can effectively be described in terms of  Wiener processes at some resolution-scale, then quantum-like features would be expected to appear in such systems\cite{nottale1993,nottale2011}.  There is a number of observational evidences that this might indeed be the case for various classes of gravitational Keplerian systems\cite{nottale1997,hermann1998,nottale2000,nottale2011}. In this context, it is worth noting here that, with the substitution  $\hbar\leftrightarrow 2m\mathcal D$, the generalized de\,Broglie wavelength for a particle of velocity $\bf v$ becomes independent of the particle's mass: $\lambda_{dB}=\frac{2\mathcal D}{|{\bf v}|}$. This implies that an ensemble of particles with different masses constituting a macroscopic chaotic system in a given quantum-like state would be distributed according to the same position probability density. There is, however, a difference between this situation and the resolution-scale relativity based interpretation of standard quantum mechanics: when observed at a fine enough resolution-scale, a given particle within the set recovers a differentiable trajectory that can be described using classical mechanics. At coarse enough resolution-scale, it may be appropriate to describe the system as being in a stationary quantum-like state labeled for example by  some eigenvalues of  mass-specific energy and angular momentum or as time evolving because of being in a superposition of a few such stationary states.  However, at resolutions finer than the generalized de Broglie wavelength characterizing the quantum like state, a given constituent of the system can be found with almost any energy and angular momentum. The one particle does not hold much information about the quantum-like state which possibly describes its motion and that of other constituents of the system observed at coarser resolutions. Considering that the interaction of the one particle with rest of the system conserves energy and angular momentum for example, these quantities for the individual particle are classically entangled with those for the rest of the ensemble of particles. Without this interactions between individual particles and the rest of the system, there would not be any chaotic motion at any scale and, consequently, no effective description in terms of Wiener processes and, hence, no quantum-like description either. The scale-relativistic quantum-like state appears as the state of the entire system regarded under some sort of thermodynamic limit under which the existence of individual constituents is lost. As such, the scale-relativistic quantum-like state would not be the state of the individual constituents whose mutual interactions are responsible for the chaotic behavior falling under a quantum like description if the resolution-scale relativity principle is applicable. Then, the motions of the individual constituents can be envisioned as as many integrations of the It\^o process for the quantum-like state, the information about which is held by the entire system. As long as one does not rely on the physical existence of some sub-quantum Brownian motion for the interpretation of quantum mechanics, this apparent disconnect  between the content and the container does not arise for standard quantum mechanics.

%%------------------------------------------------------------------------------------------------------------------------
%%------------------------------------------------------------------------------------------------------------------------
\section{Summary and conclusions}\label{conclusion}
In Section \ref{sr}, we reviewed the development of the dynamics of non-differentiable paths. It proceeds from the definition of complex and resolution-scale dependent velocity and time differential operator (Equations \ref{complexv} \& \ref{covardiff}). Assuming the system can be described by a complex Lagrange function, the equation of motion was found to take the form of Newton's fundamental relation of dynamics (Equation \ref{newton}) with the usual velocity and time derivative replaced by their complex and resolution-scale dependent counterparts. Separating the real and imaginary parts of this equation, and restricting ourselves to one dimensional cases with no net motion, we obtained a Riccati equation for the imaginary part of the complex velocity (Equation \ref{nelsonriccati}). On the basis of the definition of the time-differential operator, we find that the system evolves according to an It\^o process (Equation \ref{langevin}) driven by the imaginary part of the complex velocity, solution of the Riccati equation. 

In Section \ref{ricsec}, we found that the second order linear differential equation corresponding to the Riccati equation is Schr\"odinger's equation with the substitution  $\hbar\leftrightarrow 2m\mathcal D$. We observed that the drift term in the It\^o process tends to make the particle move away from the nodes of the wave function and toward its extrema in a way compensated for by the stochastic term in the stationary regime for which the Riccati equation was obtained. More specifically, in Section \ref{fp}, we verified that the Fokker-Planck equation for the It\^o process is solved by a probability density equal to the squared modulus of the solution of the Schr\"odinger equation corresponding to the Riccati equation. 
In Section \ref{mad}, we verified that the Madelung equations, which are equivalent to Schr\"odinger's equation, also yield the same Riccati equation when applied in the stationary regime. This two way correspondence establishes the equivalence of the generalized Newton Equation \ref{newton} of dynamics and Schr\"odinger's Equation \ref{schrodinger} at least in the one dimensional stationary case. 

In Section \ref{multidim}, we postulated an expression for the complex velocity that generalized the one dimensional result to any number of dimensions. We used it to write the multidimensional It\^o process implementing the non-differentiable dynamics and verified that the corresponding multidimensional Fokker-Planck equation is satisfied by the squared modulus of the solution to the Schr\"odinger equation, not only in the stationary case but also in the time dependent case. We then commented on the fact that the confirmed expression of the complex velocity matches that obtained from the canonical momentum calculated as the gradient of the complex action written as the logarithm of the wave function. It is specifically this identity that allows to establish the equivalence of the generalized Newton Equation \ref{newton} of dynamics with Schr\"odinger's Equation \ref{schrodinger}\cite{nottale2011,MHTeh2017}.  

The hypothesis that the resolution-scale relativity principle is implemented in nature is validated by the fact that its enforcement to mechanics leads straightforwardly to a relativity principle based foundation of standard quantum mechanics prevailing at small resolution-scales. In Section \ref{chaos}, we commented on the fact that, in the scale-relativistic interpretation, the state of the system is to be identified with the entire set of dynamical paths merely sampled by an It\^o process. We then noted that if, as suggested by a number of observations, the resolution-scale relativity principle is applicable to macroscopic chaotic systems, then the interpretation of the nature of the state of the system must be different:  a single constituent can not by itself carry the information of a quantum-like state. Instead, the state of the system has to be a collective property of the ensemble of constituents whose mutual interactions are responsible for the chaotic dynamics and the quantum-like behavior under the resolution-scale relativity principle. As such, the interpretation of the state of a chaotic macroscopic system of $N$ particles may recover the same form as for one particle in standard quantum mechanics, provided it is regarded as a single self interacting particle in a $3N$ dimensional space. Inversely, the richness of the wave function of a single standard quantum mechanical particle may be considered as an infinite collection of dynamical paths each coming under a description by the same It\^o process. In these scale relativistic interpretations of standard quantum mechanics and macroscopic chaos, the difference resides in the respectively infinite and finite numbers of degrees of freedom. As a consequence, while it is very possible that the resolution-scale relativity principle applies to macroscopic chaos, it should only do so approximatively. This leaves open the question of the manifestation of the resolution-scale relativity principle in systems demonstrating dynamical chaos over a small number of degrees of freedom. 

%See Figure \ref{paperlogic}
%\begin{figure}
%\begin{center}
%\rotatebox{0}{\includegraphics[scale=0.5]{Riccati_paper_logic.pdf}}
%\end{center}
%\caption{This is a schematic of the logic of the paper. I think that is what we should outline in the conclusion. }
%\label{paperlogic}
%\end{figure}   

%%------------------------------------------------------------------------------------------------------------------------
%%------------------------------------------------------------------------------------------------------------------------
\begin{acknowledgments}
S.\,LeBohec is grateful to Yong-Shi Wu for his helpful conversation and to Dirk P\"utzfeld for his attentive reading and corrections of the manuscript. 
\end{acknowledgments}
%%------------------------------------------------------------------------------------------------------------------------
%%------------------------------------------------------------------------------------------------------------------------

\end{document}